\pdfoutput=1
\documentclass[preprint,aps,pre,amsmath,amssymb,nofootinbib]{revtex4-2}
\usepackage{graphicx}
\usepackage{bm}
\usepackage{hyperref}
\usepackage{xcolor}

\newcommand{\ave}[1]{\left\langle #1 \right\rangle}
\newcommand{\dd}{\mathrm{d}}

\begin{document}

\title{Parisi's solution of the Sherrington--Kirkpatrick model from synchronous Monte Carlo dynamics}

\author{Yoshiyuki Kabashima}
\altaffiliation[Also at ]{Trans-Scale Quantum Science Institute, The University of Tokyo, 7-3-1 Hongo, Bunkyo-ku, Tokyo 113-0033, Japan}
\email{kaba@phys.s.u-tokyo.ac.jp}
\affiliation{%
 Institute for Physics of Intelligence, The University of Tokyo,
 7-3-1 Hongo, Bunkyo-ku, Tokyo 113-0033, Japan
}%

\date{\today}

\begin{abstract}
Sompolinsky and Zippelius obtained Parisi's solution of the Sherrington--Kirkpatrick (SK) model from the relaxational dynamics of soft spins in continuous time, by assuming a hierarchy of widely separated time scales.
Here we carry out the same programme for a discrete-time dynamics in which all Ising spins are updated at once.
This dynamics, synchronous Monte Carlo (SyncMC), introduces Gaussian auxiliary fields so that the spins become conditionally independent, and it samples the Boltzmann distribution exactly for any system with symmetric pairwise couplings.
We show that, for any such system, the correlation and response functions of SyncMC obey an exact fluctuation--dissipation relation in discrete time in equilibrium.
The dynamical mean-field equations of SyncMC for the SK model contain two terms absent in continuous-time dynamics, a self-coupling of each spin and a noise correlated through the couplings; both affect only the fast relaxation.
As a result, under the same assumptions as in the Sompolinsky--Zippelius construction (widely separated time scales, and on each slow scale the same relation with a reduced ratio $x$, as expected in relaxation from random initial conditions), the long-time structure of SyncMC reproduces the Parisi equations, including the Parisi partial differential equation and the equation for the distribution of local fields.
Equilibrium simulations with replica exchange for up to $N=2048$ spins exclude the replica-symmetric distribution of the actual local fields and are consistent with the Parisi solution.
The function $x(q)$ estimated from the overlap distribution, and from the response in relaxation from random configurations for up to $N=4096$ spins, is also consistent with Parisi's.
\end{abstract}

\maketitle

\section{Introduction}
\label{sec:intro}

Parisi's solution of the Sherrington--Kirkpatrick (SK) model \cite{SK1975,Parisi1979,Parisi1980,MPV1987} describes the low-temperature phase of a mean-field spin glass through a function $x(q)$ on the interval $0\le q\le 1$.
Soon after Parisi's work, Sompolinsky and Zippelius showed that the same structure can be read from the dynamics \cite{SZ1981,Sompolinsky1981,SZ1982}.
They studied the relaxation of soft spins described by a Langevin equation in continuous time.
They assumed that the long-time relaxation splits into a hierarchy of widely separated time scales.
On each scale, the response and correlation functions are related by a fluctuation--dissipation relation with a reduced ratio.
Under this assumption, the free energy that follows from the dynamics coincides with Parisi's, with $q$ playing the role of the time scale and $x(q)$ that of the fluctuation--dissipation ratio.
Later work extended this connection to relaxation from random initial conditions \cite{CK1994}, interpreted the ratio $x$ as an effective temperature on each time scale \cite{CKP1997}, and related the Parisi construction to systems coupled to several heat baths at different temperatures \cite{Horner1984,PCS1993,DFM1994,ANS2000,AN2000,CKM2019,CCKM2021,AMM2023}.

These works use continuous-time dynamics, mostly Langevin dynamics of soft spins.
Practical simulations of Ising systems instead use discrete-time Monte Carlo dynamics.
Updating all spins at once is attractive on parallel hardware, but the simplest such scheme, the synchronous heat-bath rule (Little dynamics) \cite{Little1974}, does not sample the Boltzmann distribution of the model; its stationary distribution is that of a different function of the spins \cite{Peretto1984}.
A way to update all spins at once while keeping the Boltzmann distribution exact is to introduce Gaussian auxiliary fields, which make the spins conditionally independent \cite{MartensSutskever2010,Zhang2012}.
We call the resulting dynamics synchronous Monte Carlo (SyncMC) \cite{SyncMC}.
It applies to any system with symmetric pairwise couplings.

In this paper we ask whether the Sompolinsky--Zippelius construction carries over to SyncMC.
SyncMC differs from Langevin dynamics in three respects:
time is discrete, all spins move at once, and the noise acting on different spins is correlated through the couplings.
Each of these could in principle change the long-time structure.
We show that it does not, and we identify the two properties that make this so.

\begin{enumerate}
\item \textbf{Exact discrete-time fluctuation--dissipation relation.}
For any symmetric couplings, the response $R(t,s)$ of a spin at time $t$ to a field applied to it at time $s$ and the correlation $C(t,s)$ of the spin at times $t$ and $s$ satisfy, in equilibrium at inverse temperature $\beta$,
$R(t,s)=\beta[C(t,s+1)-C(t,s)]$ exactly (Sec.~\ref{sec:fdt}).
This follows from the structure of SyncMC as an alternation between two conditional distributions of one joint distribution.
\item \textbf{The terms specific to SyncMC drop out on long time scales.}
The dynamical mean-field theory (DMFT) of SyncMC for the SK model contains a self-coupling of each spin and an additional noise whose covariance is set by the response function (Sec.~\ref{sec:dmft}).
Both affect only the fast relaxation.
On the slow time scales the effective single-spin process has the same form as in continuous time (Sec.~\ref{sec:slow}).
\end{enumerate}

With these two properties, the assumptions of Sompolinsky and Zippelius lead to the nested recursion for the local partition function, to the Parisi partial differential equation, and to the equation for the distribution of local fields (Sec.~\ref{sec:parisi}).
Writing these equations with $q$ as the independent variable makes the dynamical meaning of each quantity explicit: $q$ labels the time scale through the value of the correlation function, and $x(q)$ is the fluctuation--dissipation ratio on that scale.
The self-consistency condition that closes these equations appears as the statement that, on the slow scales, the noise felt by a spin has the covariance of its own correlation function, which holds because the terms specific to SyncMC have dropped out.

Thus a dynamics that updates all spins at once and samples the exact Boltzmann distribution has the same long-time organisation as the dynamics used in the original theory.
In Sec.~\ref{sec:observable} we give the distribution of the actual local fields $y_i=\sum_j J_{ij}\sigma_j+H$, where $\sigma_j=\pm1$ are the spins, $J_{ij}$ the couplings and $H$ the external field, as an observable that depends only on equilibrium and compare it with SyncMC simulations.
We also test the multi-time-scale picture itself, through the relaxation from equilibrium configurations, and estimate $x(q)$ from simulations in two ways: from the overlap distribution in equilibrium and from the response in relaxation from random configurations.

We stress what is and is not new.
The multi-time-scale route to Parisi's solution and the elimination of variables coupled to baths at different temperatures are known \cite{SZ1981,Sompolinsky1981,SZ1982,CKP1997,AN2000,CKM2019,CCKM2021}.
What we add is the analysis of a discrete-time, fully parallel dynamics with correlated noise, the exact discrete-time fluctuation--dissipation relation that holds for it, and the demonstration that its specific terms do not reach the slow time scales.
Our derivation of the Parisi structure remains at the level of physical arguments, resting on the same assumptions as in Refs.~\cite{SZ1981,SZ1982}.

\section{Synchronous Monte Carlo for pairwise systems}
\label{sec:syncmc}

Consider $N$ Ising spins $\sigma_i=\pm1$ with energy
\begin{equation}
  E(\bm\sigma) = -\frac12\sum_{i\neq j}J_{ij}\sigma_i\sigma_j - \sum_i H_i\sigma_i ,
  \label{eq:energy}
\end{equation}
where $\bm\sigma=(\sigma_1,\dots,\sigma_N)$, the coupling matrix $J=(J_{ij})$ is real and symmetric with zero diagonal, and $H_i$ is the external field on spin $i$.
Choose a constant $c\ge -\lambda_{\min}(J)$, where $\lambda_{\min}(J)$ is the smallest eigenvalue of $J$, so that $A=J+cI$ is positive semidefinite; $I$ is the $N\times N$ identity matrix.
Because $\sigma_i^2=1$, $\bm\sigma^{\mathsf T}A\bm\sigma=\bm\sigma^{\mathsf T}J\bm\sigma+cN$ (${}^{\mathsf T}$ denotes the transpose), and the Boltzmann distribution at temperature $T=1/\beta$, $\pi(\bm\sigma)\propto e^{-\beta E(\bm\sigma)}$, does not depend on $c$.

Time is measured in steps, $t=0,1,2,\dots$, and one SyncMC step reads
\begin{align}
  \bm h(t) &= A\,\bm\sigma(t) + \bm H + \bm\xi(t),\qquad \bm\xi(t)\sim\mathcal N\!\left(0,\beta^{-1}A\right),
  \label{eq:syncmc-h}\\
  \Pr[\sigma_i(t+1)=\pm1\mid \bm h(t)] &= W(\pm1\mid h_i(t)),\qquad W(\sigma\mid h)=\frac{e^{\beta\sigma h}}{2\cosh\beta h},
  \label{eq:syncmc-hb}
\end{align}
with all spins updated at once and $\bm\xi(t)$ drawn independently at each step.
Here $\bm h(t)$ is the SyncMC field, $\bm H=(H_1,\dots,H_N)$, $\mathcal N(\bm\mu,\Sigma)$ is the Gaussian distribution with mean $\bm\mu$ and covariance matrix $\Sigma$, $\Pr[\,\cdot\mid\cdot\,]$ is a conditional probability, and $W$ is the heat-bath probability.
The cost per step is two matrix--vector products, and many replicas or temperatures can be updated together as matrix--matrix products.

\paragraph*{Exact sampling.}
For positive definite $A$, introduce the joint distribution
\begin{equation}
  \pi(\bm\sigma,\bm h)=\pi(\bm\sigma)\,p(\bm h\mid\bm\sigma),\qquad
  p(\bm h\mid\bm\sigma)=\mathcal N\!\left(\bm h;\,A\bm\sigma+\bm H,\,\beta^{-1}A\right),
\end{equation}
where $\mathcal N(\bm h;\bm\mu,\Sigma)$ is the density of $\mathcal N(\bm\mu,\Sigma)$ at $\bm h$.
Writing $\bm u=\bm h-\bm H$ and expanding the Gaussian exponent, the terms quadratic in $\bm\sigma$ cancel and
\begin{equation}
  \ln\pi(\bm\sigma,\bm h)=\beta\sum_i\sigma_ih_i-\frac\beta2\,\bm u^{\mathsf T}A^{-1}\bm u+\text{const}.
\end{equation}
Hence $\pi(\bm\sigma\mid\bm h)=\prod_iW(\sigma_i\mid h_i)$.
SyncMC therefore draws alternately from the two conditional distributions of $\pi(\bm\sigma,\bm h)$, which is a two-block Gibbs sampler.
The transition probability of the spins in one step, $\mathcal K(\bm\sigma\to\bm\sigma')=\int\dd\bm h\,p(\bm h\mid\bm\sigma)\,\pi(\bm\sigma'\mid\bm h)$, satisfies
\begin{equation}
  \pi(\bm\sigma)\mathcal K(\bm\sigma\to\bm\sigma')=\int\dd\bm h\,\frac{\pi(\bm\sigma,\bm h)\,\pi(\bm\sigma',\bm h)}{\pi(\bm h)},
\end{equation}
where $\pi(\bm h)=\sum_{\bm\sigma}\pi(\bm\sigma,\bm h)$ is the marginal distribution of $\bm h$;
the right-hand side is symmetric in $\bm\sigma\leftrightarrow\bm\sigma'$.
The chain is therefore reversible with respect to the Boltzmann distribution.
The semidefinite case follows as a limit.

\section{Exact fluctuation--dissipation relation in discrete time}
\label{sec:fdt}

Let the local fields depend on time, $H_i(s)$.
Since $H_i(s)$ enters the path probability only through the factor $W(\sigma_i(s+1)\mid h_i(s))$, the response of spin $i$ is
\begin{equation}
  R_i(t,s)\equiv\frac{\partial\ave{\sigma_i(t)}}{\partial H_i(s)}
  =\beta\Bigl\langle\sigma_i(t)\bigl[\sigma_i(s+1)-\tanh\beta h_i(s)\bigr]\Bigr\rangle ,
  \label{eq:response}
\end{equation}
for $t\ge s+1$, and $R_i(t,s)=0$ for $t\le s$.
Here and below, $\ave{\cdots}$ denotes the average over the stochastic dynamics for fixed $J$, that is, over the initial configuration, the auxiliary noise $\bm\xi$ and the heat-bath updates; in the stationary state it coincides with the average over the Boltzmann distribution $\pi$.
The derivative is taken at the actual fields, and $\tanh\beta h_i(s)=\sum_{\sigma=\pm1}\sigma\,W(\sigma\mid h_i(s))$ is the mean of $\sigma_i(s+1)$ given $\bm h(s)$.
This expression holds for any $J$ and any initial condition.

\paragraph*{Proposition.}
In the stationary state of SyncMC with constant fields,
\begin{equation}
  R_i(t,s)=\beta\bigl[C_i(t,s+1)-C_i(t,s)\bigr],\qquad C_i(t,s)=\ave{\sigma_i(t)\sigma_i(s)},\qquad t\ge s+1 .
  \label{eq:dfdt}
\end{equation}

\paragraph*{Proof.}
Consider the sequence $X_0=\bm\sigma(0)$, $X_1=\bm h(0)$, $X_2=\bm\sigma(1)$, $X_3=\bm h(1),\dots$ in the stationary state.
Every adjacent pair, $(\bm\sigma(u),\bm h(u))$ or $(\bm\sigma(u+1),\bm h(u))$, is distributed according to $\pi(\bm\sigma,\bm h)$, and the transitions are the conditional distributions of $\pi$.
The sequence is therefore invariant under time reversal.
Under reversal, the pair $(\bm h(s),\bm\sigma(t))$ at positions $2s+1$ and $2t$ maps onto the pair $(\bm\sigma(s),\bm h(t-1))$ at positions $2s$ and $2t-1$.
Hence, writing $\mathbb E[\,\cdot\mid\bm h]$ for the conditional expectation given $\bm h$,
\begin{equation}
  \ave{\sigma_i(t)\tanh\beta h_i(s)}=\ave{\sigma_i(s)\tanh\beta h_i(t-1)}
  =\ave{\sigma_i(s)\,\mathbb E[\sigma_i(t)\mid\bm h(t-1)]}=C_i(t,s).
\end{equation}
Inserting this into Eq.~\eqref{eq:response} gives Eq.~\eqref{eq:dfdt}. $\square$

Equation~\eqref{eq:dfdt} holds for each spin and for each realisation of $J$; it does not require the large-$N$ limit.
In the mean-field analysis below it plays the role that the continuous-time fluctuation--dissipation theorem plays in Refs.~\cite{SZ1981,SZ1982}.

\section{Dynamical mean-field theory of SyncMC for the SK model}
\label{sec:dmft}

From now on $J_{ij}$ are independent Gaussian variables with zero mean and variance $1/N$, and $H_i=H$.
The spectrum of $J$ fills $[-2,2]$ for large $N$, so $c\ge2$ is required.

\subsection{Effective single-spin process}

We represent the path probability with response fields $\hat h_i(t)$ conjugate to $h_i(t)$, average over the noise $\bm\xi$ and over $J$, and take the saddle point for $N\to\infty$; the details are given in Appendix~\ref{app:dmft}.
The couplings enter in two ways: through the usual term $-i\hat{\bm h}^{\mathsf T}J\bm\sigma$, and through the noise covariance, which produces $-\frac1{2\beta}\hat{\bm h}^{\mathsf T}J\hat{\bm h}$.
The second contribution is specific to SyncMC.
Each spin then follows the effective process, written for a single spin $\sigma(t)$ and its field $h(t)$,
\begin{align}
  h(t)&=H+c\,\sigma(t)+\sum_{s<t}R(t,s)\,\sigma(s)+\varphi(t),
  \label{eq:dmft-h}\\
  \Pr[\sigma(t+1)=\pm1\mid h(t)]&=W(\pm1\mid h(t)),
  \label{eq:dmft-hb}\\
  \ave{\varphi(t)\varphi(s)}&=C(t,s)+\frac{c}{\beta}\,\delta_{ts}+\frac1\beta\bigl[R(t,s)+R(s,t)\bigr],
  \label{eq:dmft-cov}
\end{align}
where $\varphi$ is a Gaussian process with zero mean, $\delta_{ts}$ is the Kronecker delta, $\ave{\cdots}$ now denotes the average over the effective process (over $\varphi$ and the heat-bath updates), and
\begin{equation}
  C(t,s)=\ave{\sigma(t)\sigma(s)},\qquad
  R(t,s)=\beta\Bigl\langle\sigma(t)\bigl[\sigma(s+1)-\tanh\beta h(s)\bigr]\Bigr\rangle\quad(t\ge s+1)
  \label{eq:dmft-sc}
\end{equation}
are determined self-consistently.
Compared with continuous-time dynamics of the SK model, Eq.~\eqref{eq:dmft-h} contains the self-coupling $c\,\sigma(t)$, and Eq.~\eqref{eq:dmft-cov} contains two additional noise terms: a white noise of strength $c/\beta$ and a term set by the response function, which comes from the noise being correlated through $J$.

\subsection{Check against direct simulation}

\begin{figure}[t]
  \centering
  \includegraphics{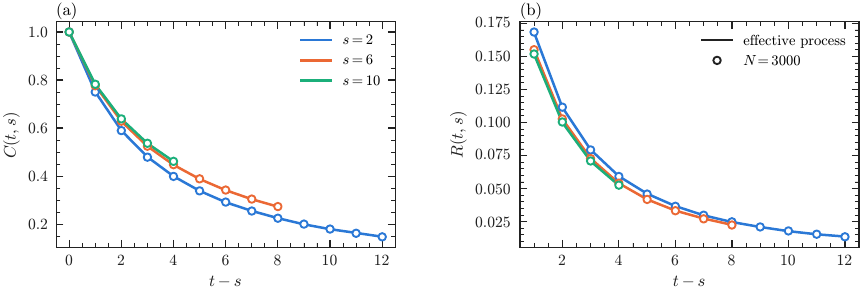}
  \caption{Correlation (a) and response (b) functions of SyncMC for the SK model at $\beta=0.7$, $H=0.3$, $c=2.3$, starting from random spins.
  Lines: solution of the effective process, Eqs.~\eqref{eq:dmft-h}--\eqref{eq:dmft-sc}, with $1.6\times10^6$ sampled paths.
  Symbols: direct simulation with $N=3000$, averaged over six realisations of $J$.
  Error bars are smaller than the symbols.}
  \label{fig:dmft}
\end{figure}

Figure~\ref{fig:dmft} compares the solution of Eqs.~\eqref{eq:dmft-h}--\eqref{eq:dmft-sc} with direct SyncMC simulations of the SK model.
We take $\beta=0.7$, $H=0.3$, $c=2.3$ and random initial spins.
The effective process is solved by sampling $1.6\times10^6$ paths with a fixed sequence of random numbers and iterating $C$ and $R$.
Because $C$ and $R$ at time $t$ depend only on their values at earlier times, the iteration converges exactly after a finite number of steps (16 here).
The simulations use $N=3000$ and six realisations of $J$.
The magnetisation, correlation and response functions agree closely: the largest differences are $0.002$ in $C$, $0.001$ in $R$ and $0.002$ in the magnetisation.
These differences are explained by the statistical error of sampling the paths of the effective process.
Solving the effective process with $4\times10^5$ paths instead gives differences about twice as large, as expected for a sampling error, and the residuals of the two solutions are not correlated with each other.
Finite-size effects at $N=3000$ are therefore not visible at the level of $10^{-3}$.
Near the end of the run, where the system is close to stationary, the data also satisfy Eq.~\eqref{eq:dfdt}: the simulation gives $R(14,12)=0.0998\pm0.0001$, the effective process gives $R(14,12)=0.0997$, and $\beta[C(14,13)-C(14,12)]=0.1004$ from the effective process.

\section{Where the terms specific to SyncMC go}
\label{sec:slow}

In the stationary state, Eq.~\eqref{eq:dfdt} states that $R(t,s)$ equals $\beta$ times the change of $C$ over one step; under assumption (ii) of Sec.~\ref{sec:parisi} the same holds with an additional factor $x\le1$.
On a time scale $\tau\gg1$ over which $C$ changes by an amount of order one, $R(t,s)$ is therefore of order $1/\tau$.
Two consequences follow.

First, the term $[R(t,s)+R(s,t)]/\beta$ in the noise covariance \eqref{eq:dmft-cov} is negligible for $|t-s|\gg1$.
Together with the white noise $c\,\delta_{ts}/\beta$, it shapes only the fast relaxation.
On the slow scales the noise covariance is $C(t,s)$, as in continuous time.

Second, the reaction term $\sum_{s<t}R(t,s)\sigma(s)$ collects contributions of order $1/\tau$ from $O(\tau)$ past times on each slow scale.
Its slow part is therefore of order one and has the same form as the continuous-time reaction term, with $\sigma(s)$ replaced by its average over faster fluctuations.
The self-coupling $c\,\sigma(t)$ acts at equal times and changes only the short-time dynamics.

Thus $c$ and the correlated noise determine how fast equilibrium within a state is reached, but they do not appear in the slow part of the effective process.
This is the second property announced in the introduction.

\section{Multi-time-scale structure and the Parisi equations}
\label{sec:parisi}

\subsection{Assumptions}

Following Refs.~\cite{SZ1981,Sompolinsky1981,SZ1982}, we assume:
\begin{itemize}
\item[(i)] \emph{Separation of time scales.}
The correlation function is a sum of components that decay on widely separated time scales $1\sim\tau_K\ll\tau_{K-1}\ll\dots\ll\tau_0$:
\begin{equation}
  C(t,s)=q_0+\sum_{k=0}^{K}\Delta_k(t-s),\qquad
  \Delta_k(\tau)=(q_{k+1}-q_k)\,c_k(\tau/\tau_k),
  \label{eq:Cdecomp}
\end{equation}
where each $c_k$ decreases from $c_k(0)=1$ to $c_k(\infty)=0$, and $0=q_0\le q_1\le\dots\le q_K=q_{\max}\le q_{K+1}=1$.
For a time difference $\tau_{k+1}\ll t-s\ll\tau_k$ the components $j>k$ have decayed and those with $j\le k$ have not, so $C(t,s)=q_{k+1}$; as $t-s$ passes through $\tau_k$, $C$ decreases from $q_{k+1}$ to $q_k$.
Correspondingly, the slow part of the Gaussian field $\varphi$ is a sum of independent components $\varphi_k$ with covariance $\Delta_k(t-s)$ (Sec.~\ref{sec:slow}).
When the components that vary on scales slower than $\tau_k$ are held fixed, the faster degrees of freedom reach a stationary state.
\item[(ii)] \emph{Fluctuation--dissipation relation on each scale.}
The response function is the sum of the responses on the individual scales,
\begin{equation}
  R(t,s)=\sum_{k=0}^{K}R_k(t,s),\qquad
  R_k(t,s)=\beta x_k\bigl[\Delta_k(t-s-1)-\Delta_k(t-s)\bigr],
  \label{eq:Rdecomp}
\end{equation}
with $0\le x_0\le x_1\le\dots\le x_K=1$.
For a time difference on scale $\tau_k$ only $\Delta_k$ changes appreciably, so this reads $R(t,s)=\beta x_k[C(t,s+1)-C(t,s)]$ there.
\end{itemize}
Assumption (ii) generalises the exact relation~\eqref{eq:dfdt}, which enters it in three ways.
First, on the fastest scale ($x_K=1$) it is Eq.~\eqref{eq:dfdt} itself and holds exactly; only the reduced ratios $x_k<1$ on the slower scales are assumed.
Second, Eq.~\eqref{eq:dfdt} fixes the discrete-time form of the assumption: the response is $\beta x_k$ times the change of the correlation over one step, rather than a time derivative, and this form makes the sum in Sec.~\ref{sec:nested} telescope.
Third, because Eq.~\eqref{eq:dfdt} is exact in equilibrium, it restricts the situations in which $x_k<1$ can occur.
In a stationary state $C$ and $R$ depend only on $t-s$.
A ratio $x_k<1$, however, requires the non-stationary situation discussed below, in which only the fastest component ($C>q_{\max}$) depends on $t-s$ alone, while the slower components depend on both times through scales that grow with $s$.
We write all components as functions of $t-s$ to keep the notation simple; the derivation uses only the decomposition into components and the ratio $x_k$ on each.

These assumptions do not describe SyncMC in equilibrium at finite $N$.
There, Eq.~\eqref{eq:dfdt} fixes the ratio to $x=1$ at all times, and this remains true for $N\to\infty$ at fixed times.
A ratio $x_k<1$ can appear only when $N\to\infty$ is taken before the system has time to move between states, so that the slow components of the field never reach equilibrium among themselves.
This is the situation of relaxation from a random initial configuration, where the ratio measured on long time scales in the SK model is expected to equal Parisi's $x(q)$ \cite{CK1994,FMPP1998}; we test this with SyncMC in Sec.~\ref{sec:xq}.
We therefore read (i) and (ii) as a description of the long-time dynamics of the infinite system.
The nested structure derived below is static and coincides with Parisi's; the equilibrium quantities compared with simulations in Sec.~\ref{sec:observable} do not depend on how it is reached.

\subsection{Nested recursion}
\label{sec:nested}

By Sec.~\ref{sec:slow}, on scale $\tau_k$ the part of the local field that varies on that scale can be written as
\begin{equation}
  u_k(t)=\varphi_k(t)+\sum_{s<t}R_k(t,s)\,\bar\sigma(s),
  \label{eq:uk}
\end{equation}
where $\varphi_k$ is the Gaussian component with covariance $\Delta_k(t-s)$, $R_k$ is the response on that scale, Eq.~\eqref{eq:Rdecomp}, and $\bar\sigma$ is the spin averaged over faster fluctuations.
This is the output of a linear Gaussian system driven by $\bar\sigma$.
Its free fluctuations have variance $\Delta_k(0)=q_{k+1}-q_k$.
If the input $\bar\sigma$ stays constant over the time $\tau_k$, the sum in Eq.~\eqref{eq:uk} telescopes:
\begin{equation}
  \sum_{s<t}R_k(t,s)\,\bar\sigma
  =\beta x_k\bar\sigma\sum_{n=1}^{\infty}\bigl[\Delta_k(n-1)-\Delta_k(n)\bigr]
  =\beta x_k\bigl[\Delta_k(0)-\Delta_k(\infty)\bigr]\bar\sigma
  =\beta x_k(q_{k+1}-q_k)\,\bar\sigma ,
\end{equation}
so a constant input shifts the mean of $u_k$ by $\beta x_k(q_{k+1}-q_k)\bar\sigma$.
These are exactly the properties of a variable with energy $u^2/2a_k-u\bar\sigma$ in equilibrium at inverse temperature $\beta_k$, with $a_k=\beta x_k(q_{k+1}-q_k)$ and $\beta_k=\beta x_k$.
In other words, the ratio $x_k$ acts as an effective temperature $T/x_k$ for the field component on scale $\tau_k$ \cite{CKP1997}.

The problem has thus become one of variables on widely separated time scales, each in contact with its own bath.
For such systems the slow variables move in the free energy of the fast ones, and their stationary distribution is a Boltzmann distribution at their own temperature \cite{AN2000,CKM2019,CCKM2021,AMM2023}.
Eliminating the variables from the fastest one upwards (Appendix~\ref{app:nested}), the field components $z_k\equiv u_k$ become Gaussian variables with variance $q_{k+1}-q_k$ reweighted by the faster partition function, and the local partition function obeys
\begin{equation}
  Z_{K+1}(y)=2\cosh\beta y,\qquad
  Z_k(y)=\Bigl\{\mathbb E_{z_k}\bigl[Z_{k+1}(y+z_k)^{x_k}\bigr]\Bigr\}^{1/x_k},
  \label{eq:nested}
\end{equation}
which is Parisi's recursion \cite{Parisi1980,MPV1987}.
On the fastest scale ($x_K=1$) the spins are in equilibrium within a state, and $Z_{K+1}$ is the partition function of a spin in its actual local field $y$.
Neither $c$ nor the correlated noise appears in Eq.~\eqref{eq:nested}.

\subsection{Continuum limit and the field distribution}

Let $q_{k+1}-q_k=\dd q\to0$, so that $x_k$ becomes a nondecreasing function $x(q)$, and let $\phi(q,h)=\beta^{-1}\ln Z_k(h)$ for $q=q_k$.
Integrating the layer $q_{\max}<q\le1$ with $x=1$ gives $Z=2\cosh\beta h\,e^{\beta^2(1-q)/2}$, so we may place the boundary at $q_{\max}$.
Expanding Eq.~\eqref{eq:nested} to first order in $\dd q$ gives the Parisi equation
\begin{equation}
  \partial_q\phi=-\frac12\bigl[\partial_h^2\phi+\beta x(q)(\partial_h\phi)^2\bigr],\qquad
  \phi(q_{\max},h)=\beta^{-1}\ln2\cosh\beta h ,
  \label{eq:parisi}
\end{equation}
and the local magnetisation $m=\partial_h\phi$ obeys
\begin{equation}
  \partial_qm=-\frac12\partial_h^2m-\beta x(q)\,m\,\partial_hm,\qquad m(q_{\max},h)=\tanh\beta h .
  \label{eq:m}
\end{equation}
The distribution $P(q,h)$ of the field seen on scale $q$ follows from adding a Gaussian increment and reweighting by $e^{\beta x\phi}$:
\begin{equation}
  \partial_qP=\frac12\partial_h^2P-\beta x(q)\,\partial_h\bigl[m\,P\bigr],\qquad P(0,h)=\delta(h-H).
  \label{eq:fp}
\end{equation}
This is the equation of Sommers and Dupont \cite{SommersDupont1984}, written with $q$ instead of $x$ as the independent variable.
The two forms are related by $\dd q=\dot q(x)\,\dd x$, where $q(x)$ is the inverse function of $x(q)$ and $\dot q=\dd q/\dd x$.
Where $q(x)$ is flat the $x$-form is trivial, and where $x(q)$ jumps the $q$-form only has a piecewise-constant coefficient; the $q$-form therefore covers both cases without change.
In the dynamical picture $q$ labels the time scale through the value of the correlation function, so it is the natural ``time'' of Eq.~\eqref{eq:fp}, while $x$ is the fluctuation--dissipation ratio on that scale, as in the representation of the fluctuation--dissipation ratio as a function $X(C)$ of the correlation in Ref.~\cite{CK1994}.

\subsection{Self-consistency condition}

On the slow scales the noise covariance of the effective process equals $C$ (Sec.~\ref{sec:slow}).
The persistent part of the correlation on scale $q$ is the mean squared magnetisation averaged over faster fluctuations, so
\begin{equation}
  q=\ave{m^2}_q\equiv\int\dd h\,P(q,h)\,m(q,h)^2,\qquad q_{\min}\le q\le q_{\max}.
  \label{eq:closure}
\end{equation}
Here $\ave{\cdots}_q$ denotes the average over $h$ with weight $P(q,h)$, and $q_{\min}$ is the smallest value of $q$ at which $x(q)>0$ ($q_{\min}=0$ for $H=0$).
This is the stationarity condition of Parisi's free energy \cite{Parisi1980,MPV1987}.
Here it follows from the dynamics, and it holds in the same form as in continuous time because the terms specific to SyncMC do not reach the slow scales.

Differentiating Eq.~\eqref{eq:closure} along the continuous part gives known relations \cite{SommersDupont1984,CrisantiRizzo2002},
\begin{equation}
  \ave{(\partial_hm)^2}_q=1 ,
  \label{eq:marginal}
\end{equation}
which states that the replicon eigenvalue vanishes at every $q$ of the continuous part, and
\begin{equation}
  x(q)=\frac{\ave{(\partial_h^2m)^2}_q}{2\beta\ave{(\partial_hm)^3}_q},
  \label{eq:xformula}
\end{equation}
which, in the present picture, expresses the fluctuation--dissipation ratio on each scale through the field distribution on that scale.
These relations are not needed in what follows; in Appendix~\ref{app:numerics} we use Eq.~\eqref{eq:xformula} only to check the numerical solution of the Parisi equations.

\paragraph*{Comparison with one-step models.}
In the SK model the equilibrium Parisi solution satisfies Eq.~\eqref{eq:marginal} on its whole continuous part, so the dynamical construction reproduces the equilibrium solution.
In models with one-step replica symmetry breaking, such as the random orthogonal model \cite{MPR1994,ParisiPotters1995,CDL2003,Caltagirone2012,SyncMC}, the time scales are discrete; the analogous marginality condition then selects marginal states rather than the equilibrium one-step solution.

\section{Observables for simulation}
\label{sec:observable}

\subsection{Distribution of the actual local field}

The quantity that SyncMC simulations measure most easily is the actual local field
\begin{equation}
  y_i=\sum_jJ_{ij}\sigma_j+H ,
\end{equation}
evaluated on equilibrium configurations.
This is not the SyncMC field \eqref{eq:syncmc-h}, which also contains $c\,\sigma_i$ and the auxiliary noise.
Extending Eq.~\eqref{eq:fp} from $q_{\max}$ to $1$ with $x=1$ gives its distribution:
\begin{equation}
  P(1,y)=\int\dd h\,P(q_{\max},h)\,
  \frac{e^{-(y-h)^2/2(1-q_{\max})}}{\sqrt{2\pi(1-q_{\max})}}\,
  \frac{2\cosh\beta y}{2\cosh\beta h}\,e^{-\beta^2(1-q_{\max})/2}.
  \label{eq:P1}
\end{equation}
Within a state, the cavity field has mean $h$, fast Gaussian fluctuations of variance $1-q_{\max}$, and the reaction of the spin enters as the weight $2\cosh\beta y$.
This expression was derived by Thomsen \textit{et al.}\ in terms of the field distribution of Sommers and Dupont [Eq.~(3.69) of Ref.~\cite{Thomsen1986}]; here it follows from extending Eq.~\eqref{eq:fp} with $x=1$.
The energy per spin follows as $E/N=-\frac12\ave{(y+H)\tanh\beta y}_{P(1,\cdot)}$, where $\ave{\cdots}_{P(1,\cdot)}$ is the average over $y$ with weight $P(1,y)$.

Equation~\eqref{eq:P1} depends only on equilibrium, so it does not depend on $c$ or on the dynamics.
The distribution of local fields over sites is the same for almost every realisation of $J$ \cite{Thomsen1986}, unlike the overlap distribution, so a single large sample averaged over sites suffices.
In the paramagnetic phase ($H=0$, $T>1$) Eq.~\eqref{eq:P1} gives $\ave{y^2}=1+\beta^2$ and $\ave{y\tanh\beta y}=\beta$; SyncMC with $\beta=0.7$, $c=2.3$ and $N=3000$ gives $1.483$ and $0.697$, respectively.

Thomsen \textit{et al.}\ compared the local-field distribution of the SK model from Monte Carlo simulations ($N=1020$) with theory, mainly through the value $P(0)$ near the transition \cite{Thomsen1986}.
Their runs were short (a few hundred Monte Carlo steps per spin) and, as they note, sample typical low-lying metastable states rather than equilibrium; the local-field distribution reached by various non-equilibrium protocols was later compared with replica theory by Horner \cite{Horner2008}.
Here we use replica exchange (exchange Monte Carlo) between temperatures \cite{HukushimaNemoto1996} to reach equilibrium, check it with an identity given below, and compare the whole distribution with both the Parisi and the replica-symmetric predictions, concentrating on the region near $y=0$ where they differ.

\begin{figure}[t]
  \centering
  \includegraphics{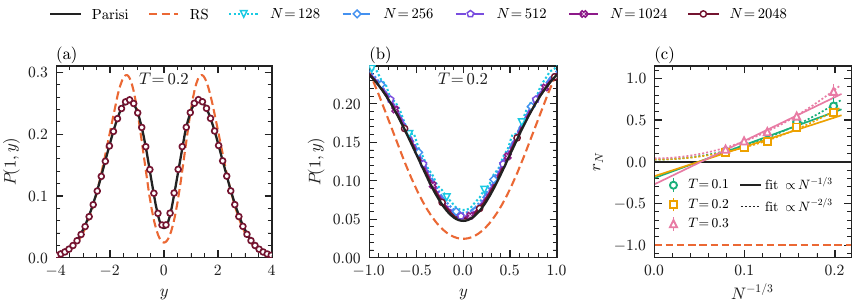}
  \caption{Distribution of the actual local field $y_i=\sum_jJ_{ij}\sigma_j$ in the SK model at $H=0$.
  (a) $P(1,y)$ at $T=0.2$: Parisi solution, replica-symmetric (RS) solution, both from Eq.~\eqref{eq:P1}, and SyncMC with replica exchange for $N=2048$ (symbols).
  (b) Enlargement near $y=0$ for all system sizes.
  (c) Finite-size dependence of the weight near $y=0$, measured by $r_N=(P_N-P_{\rm Parisi})/(P_{\rm Parisi}-P_{\rm RS})$, where each $P$ is averaged over $|y|<0.1$ with the same bins for theory and simulation.
  By construction $r_N=0$ for the Parisi solution and $r_N=-1$ for the RS solution at every temperature.
  The solid and dashed horizontal lines mark these two values.
  Symbols are plotted against $N^{-1/3}$; solid and dotted lines are fits linear in $N^{-1/3}$ and in $N^{-2/3}$, respectively, at $T=0.1$, $0.2$ and $0.3$.
  System sizes are $N=128$, 256, 512, 1024 and 2048; each size has its own colour, line style and symbol, which are the same in Figs.~\ref{fig:plateau} and \ref{fig:xqsim}.}
  \label{fig:p1}
\end{figure}

Figure~\ref{fig:p1} compares $P(1,y)$ measured with SyncMC and replica exchange with Eq.~\eqref{eq:P1} at low temperatures, for $N=128$ to $2048$.
The simulations combine SyncMC with exchanges between neighbouring temperatures at every step (even and odd pairs alternately), use two independent copies $a=1,2$ at each temperature, whose overlap is $q=N^{-1}\sum_i\sigma_i^1\sigma_i^2$, measure every 10 steps, and take $c=-\lambda_{\min}(J)+10^{-6}$; they are run in single precision on a GPU.
The temperature ladder always contains $T=0.1$, $0.2$, $0.3$, $0.4$, $0.5$ and $1.6$, with geometric spacing within each interval and a number of temperatures that grows as $\sqrt N$, from 17 for $N=128$ to 57 for $N=2048$; the smallest exchange acceptance is 0.20--0.24.
After $5\times10^4$ ($N\le256$), $2\times10^5$ ($N=512$), $3\times10^6$ ($N=1024$) and $3.5\times10^7$ ($N=2048$) steps of equilibration, we measure for $10^5$, $2\times10^5$, $10^6$ and $5\times10^6$ steps, respectively.
The numbers of samples (realisations of $J$) are 80, 80, 64, 80 and 16 for $N=128$, 256, 512, 1024 and 2048.
The equilibration identity given below is satisfied within $0.9\sigma$ for all $N\le1024$ and temperatures, and within $0.2$--$1.2\sigma$ at $T\le0.4$ and $1.8\sigma$ at $T=0.5$ for $N=2048$; block averages of the energy show no drift.
The sample-to-sample fluctuations of $\ave{q^2}$ are large (from 0.19 to 0.78 at $T=0.3$ for $N=1024$), as expected for a quantity that does not self-average, and they set the size of the errors.

The simulated distributions follow the Parisi prediction over the whole range of $y$ [Fig.~\ref{fig:p1}(a)].
The two theoretical predictions differ most near $y=0$, where the replica-symmetric solution has a deeper minimum, and this is where the finite-size effects are largest [Fig.~\ref{fig:p1}(b)].
To quantify this region we use
\begin{equation}
  r_N=\frac{P_N-P_{\rm Parisi}}{P_{\rm Parisi}-P_{\rm RS}},
\end{equation}
where $P_N$ is the measured distribution for size $N$, $P_{\rm Parisi}$ and $P_{\rm RS}$ are the two predictions, and each quantity is $P(1,y)$ averaged over $|y|<0.1$; $r_N=0$ corresponds to the Parisi solution and $r_N=-1$ to the replica-symmetric one.
At $T=0.1$, $0.2$ and $0.3$, $r_N$ is positive and decreases monotonically with $N$, from $0.59$--$0.85$ at $N=128$ to $0.11$--$0.15$ at $N=2048$ [Fig.~\ref{fig:p1}(c)].

The form of the finite-size correction to $P(1,0)$ is not known, so we compare two forms.
A correction linear in $N^{-2/3}$, the form of the correction to the energy, gives $r_\infty=0.026\pm0.029$, $0.026\pm0.020$ and $0.042\pm0.025$ at $T=0.1$, $0.2$ and $0.3$, with $\chi^2=1.2$, $0.9$ and $0.7$ for 3 degrees of freedom; these agree with the Parisi value $r_\infty=0$ within $0.9$--$1.7\sigma$.
A correction linear in $N^{-1/3}$ gives $r_\infty=-0.19\pm0.06$, $-0.17\pm0.04$ and $-0.27\pm0.05$, with $\chi^2=2.5$, $2.9$ and $1.6$; this form extrapolates past the Parisi value, by $3$--$5\sigma$.
Both forms describe the data, so the data alone do not fix the form of the correction.
With either form the limit is far from the replica-symmetric value $r=-1$.
The data therefore exclude the replica-symmetric distribution of local fields and are consistent with the Parisi solution; with the $N^{-2/3}$ form they agree with it quantitatively.

At temperatures $0.5\lesssim T\lesssim0.75$ the replica-symmetric and Parisi predictions for $P(1,y)$ differ only slightly, mainly near $y=0$; the comparison is therefore most informative at lower temperatures.
Because SyncMC at low temperature flips only a small fraction of spins per step, we combine it with replica exchange between temperatures, which is exact because the SyncMC state consists of the spins alone.
For these equilibrium measurements, sequential single-spin updates with replica exchange equilibrate faster than SyncMC (Appendix~\ref{app:numerics}); we use SyncMC because it is the dynamics analysed here.
Equilibration can be checked with the identity $\overline{\ave{E}}/N=-\frac\beta2\bigl(1-\overline{\ave{q^2}}\bigr)$ for the SK model at $H=0$, where the overline denotes the average over realisations of $J$; it follows from integration by parts over $J$.

\subsection{Time-averaged magnetisations and the plateau}
\label{sec:clones}

A second test probes the multi-time-scale picture directly.
Starting from an equilibrium configuration $\bm\sigma(0)$, we run several copies (clones) $a=1,2,\dots$ of SyncMC without temperature exchange, which differ only in their random numbers.
The overlap of two clones after $\tau$ steps,
\begin{equation}
  q_{\rm cl}(\tau)=\frac1N\sum_i\ave{\sigma_i^a(\tau)\,\sigma_i^b(\tau)},\qquad a\neq b,
\end{equation}
where the average is over the random numbers of the clones and over the equilibrium starting configuration,
and it is the mean squared magnetisation averaged over the fluctuations that are faster than $\tau$.
Unlike a time average over a finite window, it is not broadened by the finite number of samples in the window.
Because SyncMC is reversible with respect to the Boltzmann distribution (Sec.~\ref{sec:syncmc}), $\pi(\bm\sigma_0)\mathcal K^\tau(\bm\sigma_0\to\bm\sigma)=\pi(\bm\sigma)\mathcal K^\tau(\bm\sigma\to\bm\sigma_0)$, where $\mathcal K^\tau$ is the $\tau$-step transition probability,, and summing over $\bm\sigma_0$ gives
\begin{equation}
  q_{\rm cl}(\tau)=C(2\tau)
  \label{eq:qcl}
\end{equation}
exactly in equilibrium; this relation holds for any reversible dynamics.
The clone overlap therefore carries the same information as the equilibrium correlation function, and the time scale probed by a window of length $\tau$ is $2\tau$.

The separation of time scales, assumption (i), predicts how the equilibrium correlation $C$ behaves as $N$ grows.
The decay from 1 to $q_{\max}$ takes place on time scales that do not depend on $N$, while the decay below $q_{\max}$ takes place on time scales that grow with $N$.
At a fixed long time, $C$ should therefore approach $q_{\max}$ from below as $N$ increases.
For dynamics started from an equilibrium configuration, the value at which the correlation stops decaying within a state is the self-overlap of that state; Barrat \textit{et al.}\ showed this for the spherical $p$-spin model by solving the dynamical equations with an equilibrium initial condition \cite{BBM1996}.
Monte Carlo studies of the equilibrium relaxation of the SK model with single-spin updates have found time scales that grow with $N$, with $\ln\tau$ proportional to a power of $N$ \cite{MackenzieYoung1982,BillioreMarinari2001}; the present test applies the same idea to SyncMC and uses the height of the plateau to separate the Parisi and replica-symmetric predictions.
Because $x(q)$ is continuous in the SK model, there is no gap between the fast and the slow time scales, and a flat plateau is not expected at finite $N$; the test is the approach to $q_{\max}$ with increasing $N$.
The plateau height is also the quantity that separates the Parisi and replica-symmetric pictures: $q_{\max}=0.64$ versus $q_{\rm RS}=0.53$ at $T=0.5$, and $0.86$ versus $0.73$ at $T=0.3$.
Higher moments of the magnetisation within a state, such as $\ave{m^4}$, are much less useful: at the same value of $\ave{m^2}$ the two predictions differ by at most $5\times10^{-3}$ (about 1\%), which is comparable to the deviations of the data from either curve.

\begin{figure}[t]
  \centering
  \includegraphics{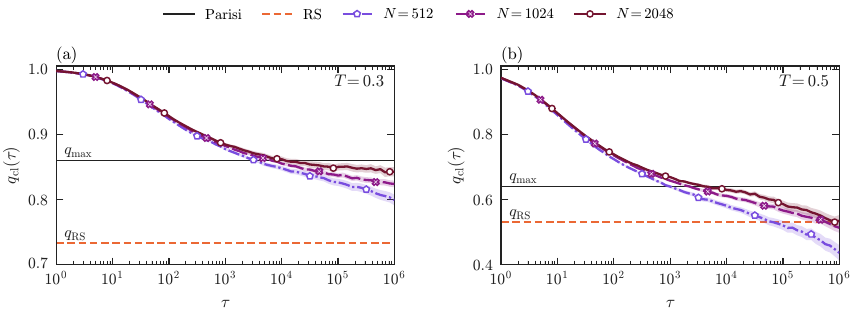}
  \caption{Overlap of two clones started from the same equilibrium configuration, $q_{\rm cl}(\tau)=C(2\tau)$, as a function of the clone time $\tau$, at $T=0.3$ (a) and $T=0.5$ (b), for $N=512$, 1024 (64 samples each) and 2048 (16 samples).
  Shaded bands are standard errors over samples.
  Horizontal lines mark the Parisi plateau $q_{\max}$ (solid) and the replica-symmetric value $q_{\rm RS}$ (dashed).}
  \label{fig:plateau}
\end{figure}

We run 16 clones started from 4 equilibrium configurations in each sample, for $\tau\le10^6$, with 64 samples for $N=512$ and $1024$ and 16 samples for $N=2048$.
The results support the prediction (Fig.~\ref{fig:plateau}).
The decay of $q_{\rm cl}(\tau)$ down to $q_{\max}$ hardly depends on $N$, and $q_{\rm cl}$ crosses $q_{\max}$ at $\tau$ of order $2\times10^3$ ($T=0.5$) and $5\times10^3$ ($T=0.3$); here $\tau$ is the time of each clone, which corresponds to $2\tau$ in $C$.
Below $q_{\max}$ the larger systems decay more slowly, and at every $\tau$ the value of $q_{\rm cl}$ increases with $N$ towards $q_{\max}$.
At $\tau=10^6$, $q_{\rm cl}=0.437\pm0.024$, $0.513\pm0.016$ and $0.535$ for $N=512$, 1024 and 2048 at $T=0.5$, and $0.799\pm0.008$, $0.824\pm0.005$ and $0.843$ at $T=0.3$.
The difference between $N=512$ and $N=1024$ is $2.7\sigma$ at $T=0.5$ and $2.5\sigma$ at $T=0.3$; the errors are dominated by sample-to-sample fluctuations, and values at different $\tau$ are strongly correlated, so the fact that the difference has the same sign at all $\tau$ is not independent evidence.
The curves for different $N$ separate near $q_{\max}$, not near $q_{\rm RS}$; in a replica-symmetric picture the decay down to $q_{\rm RS}$ would not depend on $N$.

\subsection{The ratio $x(q)$ from simulations}
\label{sec:xq}

\begin{figure}[t]
  \centering
  \includegraphics{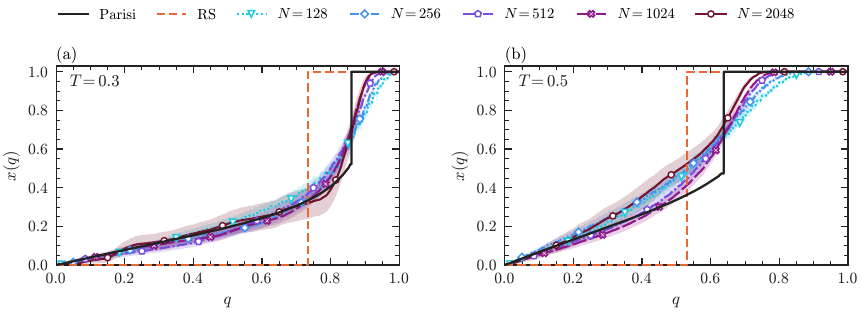}
  \caption{Function $x(q)$ estimated from the overlap distribution, Eq.~\eqref{eq:xstatic}, at $T=0.3$ (a) and $T=0.5$ (b), for $N=128$ to 2048.
  Shaded bands are standard errors over samples (64 samples for $N\le1024$, 16 for $N=2048$).
  Solid line: Parisi solution (the same as in Fig.~\ref{fig:xq}), which jumps to 1 at $q_{\max}$; dashed line: replica-symmetric solution, a step from 0 to 1 at $q_{\rm RS}$.}
  \label{fig:xqsim}
\end{figure}

The function $x(q)$ can be estimated from equilibrium simulations through the distribution of the overlap $q_{ab}=N^{-1}\sum_i\sigma_i^a\sigma_i^b$ between two independent copies at the same temperature.
In the Parisi solution $x(q)$ is the probability that $|q_{ab}|\le q$ (for $H=0$),
\begin{equation}
  x(q)=\int_{-q}^{q}P(q')\,\dd q' ,
  \label{eq:xstatic}
\end{equation}
where $P(q')$ is the distribution of $q_{ab}$ in equilibrium, averaged over $J$ (not to be confused with the field distribution $P(q,h)$),
and the estimate $x_N(q)$ for each $N$ is the fraction of measurements with $|q_{ab}|\le q$, averaged over samples \cite{Young1983}.
Because the overlap distribution differs strongly from sample to sample, many samples are needed, and the peak at $q_{\max}$ is broadened at finite $N$.
Figure~\ref{fig:xqsim} shows $x_N(q)$ obtained from equilibrium configurations saved during the replica-exchange runs.
For each sample, the overlaps between all pairs of saved snapshots of the two independent copies ($16\times16$ pairs) give an unbiased estimate of the sample's overlap distribution; the moments $\ave{q^2}$ and $\ave{|q|}$ obtained in this way agree with those measured during the runs.
We use 64 samples for $N=128$ to 1024 and 16 samples for $N=2048$.
For $q$ well below $q_{\max}$, $x_N(q)$ agrees with the Parisi $x(q)$ within about two standard errors and is clearly nonzero, whereas the replica-symmetric solution gives $x=0$ for $q<q_{\rm RS}$.
For example, at $T=0.3$ and $q=0.2$ the Parisi value is $0.078$, and $x_N=0.070\pm0.014$, $0.078\pm0.018$ and $0.087\pm0.035$ for $N=128$, 1024 and 2048.
Near $q_{\max}$ the jump of the Parisi $x(q)$ to 1 is rounded at finite $N$: at $q=q_{\max}-0.05$, $x_N$ exceeds the Parisi value by up to 0.19, without a clear trend over the present range of $N$.
Resolving the jump requires larger systems.

The dynamical definition of $x$ as a fluctuation--dissipation ratio cannot be used in equilibrium, where Eq.~\eqref{eq:dfdt} gives $x=1$ exactly.
In relaxation from a random configuration, the integrated response $\chi(t,t_w)=\sum_{s=t_w}^{t-1}R(t,s)$, with $R(t,s)=N^{-1}\sum_iR_i(t,s)$, to a field switched on at the waiting time $t_w$, plotted against $C(t,t_w)$, has slope $-\beta X$ on long time scales, where $X$ is the fluctuation--dissipation ratio, and $X(C)$ is expected to coincide with $x(q)$ in the SK model \cite{CK1994,FMPP1998}.
With SyncMC the response can be measured without applying a field, through Eq.~\eqref{eq:response}, which holds for any initial condition; here we instead measure the integrated response directly by applying a small field.
Starting from a random configuration, we run pairs of SyncMC chains with the same initial configuration and the same random numbers up to the waiting time $t_w$, and from $t_w$ on apply the fields $+\epsilon\eta_i$ and $-\epsilon\eta_i$ to the two chains, with random signs $\eta_i=\pm1$.
With $\sigma_i^{\pm}(t)$ the spins of the chains with the fields $\pm\epsilon\eta_i$, $\chi(t,t_w)=(2\epsilon N)^{-1}\sum_i\eta_i\ave{\sigma_i^+(t)-\sigma_i^-(t)}$ up to corrections of order $\epsilon^2$, which we remove by combining $\epsilon=0.03$ and $0.06$ as $[4\chi(\epsilon)-\chi(2\epsilon)]/3$; $C(t,t_w)$ is averaged over the two chains.
We use $N=2048$ at $T=0.5$, 64 samples with 64 pairs of chains each, and $t_w=10^3$, $10^4$ and $10^5$, with $t-t_w$ up to about $10^5$ for the first two and $10^6$ for the last.

\begin{figure}[t]
  \centering
  \includegraphics{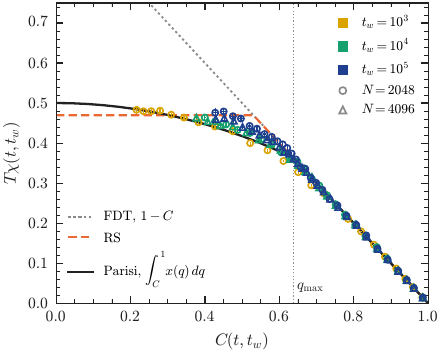}
  \caption{Integrated response $T\chi(t,t_w)$ against the correlation $C(t,t_w)$ in relaxation from random configurations, at $T=0.5$.
  Circles: $N=2048$ (64 samples) for $t_w=10^3$, $10^4$ and $10^5$; triangles: $N=4096$ (32 samples) for $t_w=10^4$ and $10^5$.
  Colours distinguish the waiting times.
  For clarity every second time point is shown.
  Solid line: $\int_C^1x(q)\,\dd q$ with the Parisi $x(q)$ of Fig.~\ref{fig:xq} ($x=1$ for $q>q_{\max}$).
  Dotted line: fluctuation--dissipation relation, $1-C$.
  Dashed line: replica-symmetric prediction, $1-C$ for $C>q_{\rm RS}$ and $1-q_{\rm RS}$ below.
  The vertical dotted line marks $q_{\max}$.
  Statistical errors are at most 0.011, comparable to the symbol size.}
  \label{fig:aging}
\end{figure}

Figure~\ref{fig:aging} shows $T\chi$ against $C$.
For $C>q_{\max}$ the data follow the fluctuation--dissipation line $T\chi=1-C$.
For $C<q_{\max}$ they leave this line and follow the curve $T\chi=\int_C^1x(q)\,\dd q$ obtained from the Parisi $x(q)$, while the fluctuation--dissipation line lies 0.15--0.3 higher at the smallest $C$ reached.
For $t_w\le10^4$ the data lie within 0.02 of the Parisi curve; the statistical errors are 0.003--0.008.
We also ran $N=4096$ (32 samples) at $t_w=10^4$ and $10^5$ (triangles in Fig.~\ref{fig:aging}); at $t_w=10^4$ the deviations from the Parisi curve are the same as for $N=2048$ within errors.
At $t_w=10^5$ the data lie above the Parisi curve, towards the fluctuation--dissipation line, by 0.02--0.05 for $N=2048$ at $0.42\le C\le0.6$.
For $N=4096$ at the same $t_w$ the deviation is smaller, 0.01--0.03; at $C=0.5$ it is $0.023\pm0.006$, compared with $0.040\pm0.008$ for $N=2048$.
This difference is $1.4$--$1.7\sigma$, so part of the deviation at $t_w=10^5$ may be a finite-size effect, but the data do not establish this.
Even for $N=4096$ the deviation at $t_w=10^5$ is larger than at $t_w=10^4$; how the deviation depends on $t_w$ is not resolved by the present data.
The replica-symmetric solution predicts that the fluctuation--dissipation line holds down to $C=q_{\rm RS}$ and that $T\chi$ is constant, $1-q_{\rm RS}$, below; the data leave the fluctuation--dissipation line near $q_{\max}$ instead, and bend smoothly without a kink at $q_{\rm RS}$.

\section{Discussion}
\label{sec:discussion}

We have shown that the Sompolinsky--Zippelius construction of Parisi's solution carries over to a discrete-time dynamics in which all spins are updated at once.
Two properties make this possible: the exact discrete-time fluctuation--dissipation relation~\eqref{eq:dfdt}, which holds for any symmetric pairwise couplings, and the confinement of the terms specific to SyncMC to the fast relaxation.

\paragraph*{Relation to earlier work.}
The multi-time-scale construction and the identification of $x$ with an effective temperature are those of Refs.~\cite{SZ1981,Sompolinsky1981,SZ1982,CKP1997}.
The step from variables at different temperatures to Parisi's nested structure has been studied for systems coupled to several baths \cite{Horner1984,PCS1993,DFM1994,ANS2000,AN2000,CKM2019,CCKM2021}, and its convergence has been analysed rigorously for two temperatures and soft spins \cite{AMM2023}.
In those works the slow variables and their temperatures are imposed from outside.
Here the slow variables are components of the cavity field of the original model, and their temperatures $T/x_k$ come from the fluctuation--dissipation relation on each scale.
Ref.~\cite{CCKM2021} also notes that the multibath arguments should hold for Glauber or Monte Carlo spin dynamics; our analysis makes this explicit for SyncMC, including the correlated noise.

\paragraph*{Limitations.}
Assumptions (i) and (ii) are taken over from Refs.~\cite{SZ1981,SZ1982} and are not derived here.
As discussed in Sec.~\ref{sec:parisi}, a ratio $x<1$ cannot occur in the equilibrium dynamics of a finite system, because Eq.~\eqref{eq:dfdt} is exact there; the assumptions refer to the infinite system on long time scales, which simulations reach through relaxation from random configurations \cite{CK1994}.
For the SK model the ratio measured in that regime is expected to agree with Parisi's $x(q)$ \cite{FMPP1998}; our data agree within 0.02 for $t_w\le10^4$, but the larger deviation at $t_w=10^5$ shows that finite times and finite $N$ require care (Sec.~\ref{sec:xq}).
The effective process of Sec.~\ref{sec:dmft} is derived for Gaussian couplings.
Equation~\eqref{eq:dfdt} holds for any couplings, but the form of the effective process for other ensembles, for example rotationally invariant couplings of another type such as the random orthogonal model, has to be worked out separately~\cite{SyncMC}.

\paragraph*{Outlook.}
The approach of $C$ to the plateau with increasing $N$ (Sec.~\ref{sec:clones}) tests assumptions (i) and (ii) directly; larger systems and longer times would sharpen this test, resolve the jump of $x(q)$ at $q_{\max}$ in the overlap distribution, and settle the dependence on $t_w$ of the response in relaxation from random configurations (Sec.~\ref{sec:xq}).
The same analysis applies to other pairwise models sampled with SyncMC, including models with one-step replica symmetry breaking, where the critical slowing down near the dynamical transition can be studied with the exponents of Ref.~\cite{Caltagirone2012}.

\begin{acknowledgments}
Generative-AI tools (Claude Opus 5.5 and Claude Fable 5.1, Anthropic) were used extensively, under the author's direction, for calculations, code, simulations, data analysis, and drafting the text. The author checked the derivations and results, including by comparing theory with simulation, and takes full responsibility for the content of the manuscript.
This work was supported by JSPS KAKENHI Grant Nos.~22H05117 and 26K02981, and JST, ARiSE, Japan Grant Number JPMJAR2633.
\end{acknowledgments}

\section*{Data availability statement}
The code that generates all numerical data reported in this article is openly available at \url{https://doi.org/10.5281/zenodo.23114947} under the MIT License.
The numerical data are not archived; they can be regenerated with this code, and are available from the author upon reasonable request.

\appendix

\section{Derivation of the effective process}
\label{app:dmft}

\paragraph*{Generating functional.}
With time-dependent fields $H_i(t)$, the probability of a path $\bm\sigma(0),\dots,\bm\sigma(t_{\rm f})$ up to a final time $t_{\rm f}$ is
\begin{equation}
  P[\bm\sigma]=p_0(\bm\sigma(0))\prod_{t=0}^{t_{\rm f}-1}\mathbb E_{\bm\xi(t)}\Bigl[\prod_iW\bigl(\sigma_i(t+1)\mid h_i(t)\bigr)\Bigr],
\end{equation}
where $p_0=\prod_ip_0(\sigma_i(0))$ is the distribution of the initial configuration, taken to be a product over sites, and $\mathbb E_{\bm\xi(t)}$ is the average over $\bm\xi(t)\sim\mathcal N(0,\beta^{-1}A)$; $Z[H]=\sum_{\bm\sigma}P[\bm\sigma]=1$ for any $H$.
Inserting $1=\int\frac{\dd h_i(t)\dd\hat h_i(t)}{2\pi}\exp\{i\hat h_i(t)[h_i(t)-(A\bm\sigma(t))_i-H_i(t)-\xi_i(t)]\}$ and averaging over $\bm\xi$ gives $e^{-\hat{\bm h}^{\mathsf T}A\hat{\bm h}/2\beta}$.
With $i=\sqrt{-1}$, $\mathcal Dh\,\mathcal D\hat h=\prod_{i,t}\dd h_i(t)\dd\hat h_i(t)/2\pi$, and separating $A=J+cI$,
\begin{align}
  Z&=\sum_{\bm\sigma}\int\mathcal Dh\,\mathcal D\hat h\,\prod_i\mathcal L_i\,\exp\Bigl[\sum_{i\neq j}J_{ij}X_{ij}\Bigr],\\
  \mathcal L_i&=p_0(\sigma_i(0))\prod_tW\bigl(\sigma_i(t+1)\mid h_i(t)\bigr)\nonumber\\
  &\quad\times\exp\Bigl\{\sum_t\Bigl(i\hat h_i(t)\bigl[h_i(t)-c\sigma_i(t)-H_i(t)\bigr]-\frac{c}{2\beta}\hat h_i(t)^2\Bigr)\Bigr\},\\
  X_{ij}&=\sum_t\Bigl[-i\hat h_i(t)\sigma_j(t)-\frac1{2\beta}\hat h_i(t)\hat h_j(t)\Bigr].
\end{align}

\paragraph*{Average over couplings.}
With $Y_{ij}=X_{ij}+X_{ji}$ and $J_{ij}\sim\mathcal N(0,1/N)$, the average over $J$ (denoted by the overline) is $\overline{\exp[\sum_{i<j}J_{ij}Y_{ij}]}=\exp[\frac1{4N}\sum_{i,j}Y_{ij}^2+O(1)]$.
Introducing $C(t,s)=\frac1N\sum_i\sigma_i(t)\sigma_i(s)$, $\hat Q(t,s)=\frac1N\sum_i\hat h_i(t)\hat h_i(s)$ and $\Gamma(t,s)=\frac1N\sum_i\hat h_i(t)\sigma_i(s)$, a direct expansion gives $\frac1{4N}\sum_{ij}Y_{ij}^2=NF$ with
\begin{equation}
  F=\sum_{t,s}\Bigl[-\frac12\hat QC-\frac12\Gamma(t,s)\Gamma(s,t)+\frac{i}{2\beta}\hat Q\bigl(\Gamma(t,s)+\Gamma(s,t)\bigr)+\frac1{4\beta^2}\hat Q^2\Bigr].
\end{equation}

\paragraph*{Saddle point.}
Imposing the definitions with conjugate variables, the sites decouple, and at the saddle point each spin carries the weight
$\mathcal L\exp\{\sum_{t,s}[\frac{\partial F}{\partial C}\sigma\sigma+\frac{\partial F}{\partial\hat Q}\hat h\hat h+\frac{\partial F}{\partial K}\hat h\sigma]\}$.
Normalisation, $Z[H]=1$, gives $\hat Q=0$, and $R(t,s)=-i\ave{\sigma(t)\hat h(s)}$ gives $\Gamma(s,t)=iR(t,s)$.
Then $\partial F/\partial C=0$, $\partial F/\partial\hat Q(t,s)=-\frac12C(t,s)-\frac1{2\beta}[R(t,s)+R(s,t)]$ and $\partial F/\partial\Gamma(t,s)=-iR(t,s)$.
Integrating over $\hat h$ yields Eqs.~\eqref{eq:dmft-h}--\eqref{eq:dmft-cov}.
Equation~\eqref{eq:dmft-sc} for $R$ follows as in Eq.~\eqref{eq:response}, since in the effective process $H(s)$ enters only the heat-bath factor at step $s$.

\section{Elimination of the slow variables}
\label{app:nested}

Fix the field components slower than scale $k$, with sum $y_{<k}=H+\sum_{\ell<k}u_\ell$.
Suppose that, for fixed $u_k$, the faster subsystem has partition function $Z_{k+1}(y_{<k}+u_k)$, free energy $F_{k+1}=-\beta^{-1}\ln Z_{k+1}$, and mean spin $\bar\sigma=m_{k+1}=-\partial_yF_{k+1}$; for $k=K$ this holds with $Z_{K+1}=2\cosh\beta y$.
Because $u_k$ is slow, the faster subsystem is stationary at each value of $u_k$, and the force on $u_k$ is $-\partial_u[u^2/2a_k+F_{k+1}(y_{<k}+u)]$.
With its bath at inverse temperature $\beta x_k$, the stationary distribution of $u_k$ is
\begin{equation}
  P(u_k\mid y_{<k})\propto e^{-u_k^2/2(q_{k+1}-q_k)}\,Z_{k+1}(y_{<k}+u_k)^{x_k},
\end{equation}
using $\beta x_k/a_k=1/(q_{k+1}-q_k)$.
The free energy of the subsystem including $u_k$, $F_k=-(\beta x_k)^{-1}\ln\mathbb E_{z_k}[Z_{k+1}(y_{<k}+z_k)^{x_k}]$, gives Eq.~\eqref{eq:nested} with $F_k=-\beta^{-1}\ln Z_k$.
Differentiating, $-\partial_yF_k$ is $m_{k+1}$ averaged over the stationary distribution of $u_k$, which closes the induction.
This elimination is the construction of Refs.~\cite{AN2000,CKM2019,CCKM2021}; see Eq.~(5) and Sec.~3.2 of Ref.~\cite{AMM2023}.

\section{Numerical methods}
\label{app:numerics}

\begin{figure}[t]
  \centering
  \includegraphics{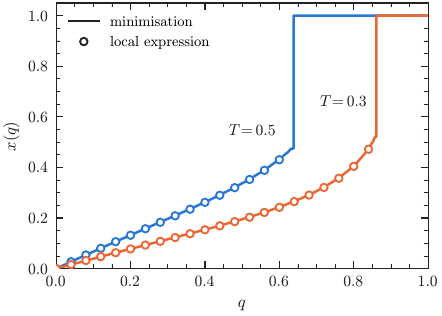}
  \caption{Fluctuation--dissipation ratio $x(q)$ of the SK model at $H=0$ for $T=0.5$ and $T=0.3$.
  Lines: direct minimisation of the Parisi functional ($M=1600$, $\Delta h=0.025$).
  Symbols: local expression~\eqref{eq:xformula} evaluated on the same solution.
  The vertical lines mark the jump to $x=1$ at $q_{\max}$; the one or two grid intervals just below $q_{\max}$, where the discretised minimiser takes values between the continuous part and 1, are replaced by the continuous part extended to $q_{\max}$ (over at most $0.003$ in $q$).}
  \label{fig:xq}
\end{figure}

\paragraph*{Parisi solution.}
We represent $x(q)$ as a nondecreasing, piecewise-constant function on a uniform grid of $M=200$ intervals in $0\le q\le1$, and compute $\phi$ by the backward recursion~\eqref{eq:nested} and $P$ by the forward reweighting, with Gaussian convolutions on a grid of spacing $\Delta h=0.025$.
We minimise the functional $-\beta f=\beta\phi(0,H)-\frac{\beta^2}2\int_0^1q\,x(q)\,\dd q$ (with $\phi(1,h)=\beta^{-1}\ln2\cosh\beta h$), which is convex in $x$, by an accelerated projected-gradient method whose gradient is proportional to $\ave{m^2}_q-q$.
The energy computed from $P(1,y)$ and from $-\frac\beta2(1-\int_0^1q(x)^2\dd x)$ agree to $10^{-4}$.
For example, at $T=0.5$ we find $q_{\max}=0.64$ and $E/N=-0.7116$, and at $T=0.2$, $f=-0.7651$ and $E/N=-0.7596$, on the two sides of the ground-state energy $-0.7633$.
As a further check, Fig.~\ref{fig:xq} compares the minimiser with Eq.~\eqref{eq:xformula} evaluated on the same solution.
With $M=200$ the minimiser forms a short plateau just below $q_{\max}$ on the discrete $q$ grid (of length $0.025$ at $T=0.5$).
For the curves of $x(q)$ in Figs.~\ref{fig:xqsim}, \ref{fig:aging} and \ref{fig:xq} we therefore use $M=1600$, for which this plateau shrinks to a single interval of width $6\times10^{-4}$; $q_{\max}$, the energy and $x(q)$ below $q_{\max}-0.05$ change by less than $2\times10^{-3}$, $10^{-5}$ and $4\times10^{-3}$, respectively.
With $M=1600$ the difference between the minimiser and Eq.~\eqref{eq:xformula} is at most $3.3\times10^{-3}$ in the interior of the continuous part (about $1\times10^{-3}$ in root-mean-square) and at most $8\times10^{-3}$ within $0.05$ of $q_{\max}$.

\paragraph*{SyncMC with replica exchange.}
The noise is generated as $\bm\xi=B\bm w/\sqrt\beta$, where $\bm w$ is a vector of independent standard Gaussian variables and $B=V\mathrm{diag}(\sqrt{\lambda+c})$, with the eigenvalues $\lambda$ and the matrix of eigenvectors $V$ of $J=V\mathrm{diag}(\lambda)V^{\mathsf T}$, and we use $c=-\lambda_{\min}(J)$.
Neighbouring inverse temperatures $\beta_a$ and $\beta_b$ exchange configurations, with energies $E_a$ and $E_b$, with probability $\min\{1,e^{(\beta_a-\beta_b)(E_a-E_b)}\}$ \cite{HukushimaNemoto1996}.
For $N=12$ the energies agree with exact enumeration within $10^{-3}$ at three temperatures.

\paragraph*{Comparison with single-spin updates.}
For reference we also ran a sequential heat-bath dynamics, which updates one spin at a time, with the same replica exchange, on the same GPU (NVIDIA H200), with the same couplings and temperature ladders, for $N=256$, 512 and 1024 and lowest temperatures $T=0.2$ and $0.3$.
Both dynamics give the same equilibrium values.
The cost of one sweep is similar for the two, but the number of sweeps needed for a replica to travel between the lowest and highest temperatures is 6--11 times smaller with sequential updates, so the computer time needed for equilibration is 6--10 times shorter; the ratio grows with $N$ and as the lowest temperature decreases.
We attribute the slower relaxation of SyncMC per step to the self-coupling $c\approx2$, which keeps the fraction of spins flipped per step small ($7\times10^{-4}$ at $T=0.3$).
We use SyncMC here because it is the dynamics analysed in this paper, not for its speed.

\end{document}